\documentclass[preprint,superscriptaddress,preprintnumbers,amsmath,amssymb,pra]{revtex4}
\usepackage{graphicx}
\usepackage{amsmath}\usepackage{slashed}
\usepackage{color}
\usepackage{mathtools}
\usepackage{txfonts}

\usepackage{xcolor}
\definecolor{darkgreen}{rgb}{0,.5,0}
\usepackage[colorlinks,filecolor=blue,citecolor=darkgreen,unicode]{hyperref}

\UseRawInputEncoding

\begin{document}

\thispagestyle{empty}

\title{Comment on ``Electric conductivity of graphene: Kubo model versus a nonlocal
quantum field theory model" }

\author{M.~Bordag}
\affiliation{Institut f\"{u}r Theoretische Physik, Universit\"{a}t Leipzig,
D-04081, Leipzig, Germany
}
\affiliation{Bogoliubov Laboratory of Theoretical Physics,
Joint Institute for Nuclear Research, 141980, Dubna, Russia}
\author{
N.~Khusnutdinov}
\affiliation{CMCC, Universidade Federal do ABC,
Avenida dos Estados 5001, CEP 09210-580, Santo Andr\'{e}, SP, Brazil}
\author{
G.~L.~Klimchitskaya}
\affiliation{Central Astronomical Observatory at Pulkovo of the Russian Academy of Sciences, St.Petersburg,
196140, Russia}
\affiliation{Peter the Great Saint Petersburg
Polytechnic University, Saint Petersburg, 195251, Russia}
\author{
V.~M.~Mostepanenko}
\affiliation{Central Astronomical Observatory at Pulkovo of the Russian Academy of Sciences, St.Petersburg,
196140, Russia}
\affiliation{Peter the Great Saint Petersburg
Polytechnic University, Saint Petersburg, 195251, Russia}

\begin{abstract}
Recently, Rodriguez-Lopez, Wang, and Antezza [Phys. Rev. B ${\bf 111}$, 115428 (2025)]
compared the theoretical descriptions of electric conductivity of graphene
given by the Kubo model and quantum field theory in terms of the polarization
tensor. According to this article, in the spatially nonlocal case, the
quantum field theoretical description contains ``hard inconsistencies".
By modifying the equality, which relates the conductivity and polarization expressions,
the predictions of quantum field theory were revised and brought in agreement with 
those following from the nonrelativistic Kubo model. Here, it is shown that
this modification violates the requirement of gauge invariance and, thus,
is unacceptable. By comparing both theoretical approaches, we demonstrate that
all the results obtained within quantum field theory are physically well
justified whereas an application of the modified expression for the
conductivity of graphene leads to the consequences of nonphysical character.
\end{abstract}

\maketitle

Reference \cite{1} compares the expressions for the electric conductivity of
graphene. One of them was obtained \cite{2,3,4,5} in the framework of Dirac
model using the formalism of quantum field theory and, specifically, the concept
of the polarization tensor \cite{6,7,8,9,10}.
The other one was found using the nonrelativistic nonlocal Kubo model \cite{1}. 
According to Ref.~\cite{1}, in the spatially nonlocal region, the transverse
conductivity of graphene obtained using quantum field theory shows a
``nonphysical plasmalike behavior\ldots~at low frequencies" leading to
``dissipation-less permanent currents".  By making a modification of the equality 
connecting the polarization and electric conductivity expressions, Ref.~\cite{1} 
revised the results obtained using relativistic quantum field theory and brought 
them in agreement with those found using the nonrelativistic nonlocal Kubo model. 

Below we demonstrate that the modified expression used in Ref.~\cite{1}
is in violation of the requirement of gauge invariance. It is shown that
the results found using this expression are physically unacceptable.
Thus, no modification is needed, and the objections of Ref.~\cite{1} against
the quantum field theoretical description are invalid.
We also list
several other inconsistencies contained in Ref.~\cite{1}.

In application to graphene, the formalism of the polarization tensor was
elaborated in Refs.~\cite{6,7,8,9} and further developed and clarified in
Refs.~\cite{10,12,13,14,15,16} (see also Ref.~\cite{17} and literature
therein for the previous publications on this subject).
In momentum representation, the components of the polarization tensor
$\Pi_{\mu\nu}$ with $\mu,\,\nu=0,\,1,\,2$ depend on the frequency $\omega$ and
on the wave vector projection $\mbox{\boldmath$q$}=(q_1,\,q_2)$ on the plane
of graphene. For a doped and gapped graphene, they also depend on the energy gap
parameter  $\Delta$, chemical potential $\mu$ and, at nonzero temperature,
on $T$.

In the absence of constant in time, external magnetic field,
the polarization tensor of graphene can be expressed via
two independent quantities, e.g.  \cite{1,2,3,4,5,13,14},
$~\Pi_{00}(\omega,\mbox{\boldmath$q$})$ and
\begin{equation}
\Pi(\omega,\mbox{\boldmath$q$})\equiv
q^2\Pi_{\mu}^{\,\mu}(\omega,\mbox{\boldmath$q$})
+\left(\frac{\omega^2}{c^2}-q^2\right)\,
\Pi_{00}(\omega,\mbox{\boldmath$q$}).
\label{eq1}
\end{equation}
\noindent
In terms of these quantities, it takes the form \cite{7}
\begin{equation}
\Pi_{\mu\nu}(\omega,\mbox{\boldmath$q$})=
\left(
\begin{array}{ccc}
\Pi_{00} & \frac{q_0q_1}{q^2}\Pi_{00}&\frac{q_0q_2}{q^2}\Pi_{00}\\[3mm]
\frac{q_0q_1}{q^2}\Pi_{00} &
\frac{q_0^2q_1^2}{q^4}\Pi_{00}-\frac{q_2^2}{q^4}\Pi &
\frac{q_1q_2}{q^4}(q_0^2\Pi_{00}+\Pi) \\[3mm]
\frac{q_0q_2}{q^2}\Pi_{00}&
\frac{q_1q_2}{q^4}(q_0^2\Pi_{00}+\Pi)&
\frac{q_0^2q_2^2}{q^4}\Pi_{00}-\frac{q_1^2}{q^4}\Pi
\end{array}\right),
\label{eq2}
\end{equation}
\noindent
where $q^2=q_1^2+q_2^2$ and $q_0=\omega/c$.

According to Ref.~\cite{1}, the tensor of electric conductivity used in
quantum field theoretical formalism of Refs.~\cite{2,3,4,5,10,12,18,19}
is given by $\sigma_{\mu\nu}=\Pi_{\mu\nu}/(-i\omega)$.
 In fact, however, Refs.~\cite{10,12,19} use this equation with
$+i\omega$ in denominator.
 As to Refs.~\cite{2,3,4,5,18}, they consider
only the longitudinal and transverse conductivities.
It is pertinent to note that Ref.~\cite{1} does not specify the used physical
units and throughout the paper presents equations written in
different systems of units (see below).
Here, we present all the following equations in the Gaussian
units. For the sake of clarity, we preserve the velocity of
light $c$, the Planck constant $\hbar$, and the Boltzmann constant $k_B$.

Under this convention, the electric conductivity tensor used in
Refs.~\cite{10,12,15,16,19} is given by
\begin{equation}
\sigma_{\mu\nu}(\omega,\mbox{\boldmath$q$})=\frac{c^2}{4\pi\hbar}\,
\frac{\Pi_{\mu\nu}(\omega,\mbox{\boldmath$q$})}{i\omega}.
\label{eq3}
\end{equation}

As discussed above, Ref.~\cite{1} considers Eq.~(\ref{eq3}) as unsatisfactory
and writes it in the modified ``regularized" form with the same values of both 
indices:
\begin{eqnarray}
\sigma_{\mu\nu}^{\rm K}(\omega,\mbox{\boldmath$q$})&=&\frac{c^2}{4\pi\hbar}\,
\frac{\widetilde{\Pi}_{\mu\nu}(\omega,\mbox{\boldmath$q$})}{i\omega}
\nonumber \\
&\equiv&
\frac{c^2}{4\pi\hbar}\,
\frac{\Pi_{\mu\nu}(\omega,\mbox{\boldmath$q$})-
\lim\limits_{\omega\to 0}\Pi_{\mu\nu}(\omega,\mbox{\boldmath$q$})}{i\omega}.
\label{eq4}
\end{eqnarray}
\noindent
Note that Eq.~(\ref{eq4}) differs from the corresponding Eqs.~(\ref{eq3}),
(93), and (95) in Ref.~\cite{1} by the constant factor which is not important
in the given context.

In Ref.~\cite{1}, Eq.~(\ref{eq4}) is derived from the Kubo formula expressing the
conductivity tensor via the current-current correlator. In this derivation, the
nonrelativistic realization of causality has been used reflected in the one-sided Fourier
transformation of all relevant quantities. In the framework of the Dirac model,
however, graphene is described by the relativistically covariant Dirac equation.
The quantum field theoretical formalism appropriate in this case employs the
Feynman Green functions, which originate from a time-ordered product of field
operators and describe the scattering of
particles in both directions in time. This corresponds to the relativistic causality
reflected in using the two-sided Fourier transformation. In the framework of relativistic
formalism appropriate for graphene described by the Dirac model, one arrives 
to Eq.~(\ref{eq3}) with no any ``regularization" \cite{19,19a,19b,19c,19d}.

{}From Eq.~(\ref{eq2}) one obtains
\begin{equation}
\lim\limits_{\omega\to 0}\Pi_{\mu\nu}(\omega,\mbox{\boldmath$q$})=
\left(
\begin{array}{ccc}
\lim\limits_{\omega\to 0}\Pi_{00} & 0 & 0\\
0 & -\frac{q_2^2}{q^4}\lim\limits_{\omega\to 0}\Pi &
\frac{q_1q_2}{q^4}\lim\limits_{\omega\to 0}\Pi \\
0 &\frac{q_1q_2}{q^4}\lim\limits_{\omega\to 0}\Pi &
-\frac{q_1^2}{q^4}\lim\limits_{\omega\to 0}\Pi
\end{array}
\right).
\label{eq5}
\end{equation}

Due to the gauge invariance, the polarization tensor $\Pi_{\mu\nu}$ satisfies the transversality condition
$q_{\mu}\Pi^ {\mu\nu}(\omega,\mbox{\boldmath$q$})=0$.
Due to Eq.~(\ref{eq3}), the same condition
$q_{\mu}\sigma^ {\mu\nu}(\omega,\mbox{\boldmath$q$})=0$
is also valid for the components of the tensor of electric conductivity.

It is easily seen, however, that the expressions
$\widetilde{\Pi}_{\mu\nu}$ and
 $\sigma_{\mu\nu}^{\rm K}$
defined in Eq.~(\ref{eq4}) do not satisfy the
transversality condition. Really, from Eq.~(\ref{eq5})
one finds that
\begin{equation}
q_{\mu}\widetilde{\Pi}^{\mu 0}(\omega,\mbox{\boldmath$q$})=
-q_{\mu}\lim\limits_{\omega\to 0}\Pi^{\mu 0}(\omega,\mbox{\boldmath$q$})
=-q_0\lim\limits_{\omega\to 0}\Pi^{0 0}(\omega,\mbox{\boldmath$q$})\neq 0
\label{eq12}
\end{equation}
\noindent
because even for the sheet of pristine graphene  ($\Delta=\mu=0$)
at fixed $q$ it holds \cite{21}
\begin{equation}
\lim\limits_{\omega\to 0}\Pi^{0 0}(\omega,\mbox{\boldmath$q$})=
\frac{\alpha\pi\hbar cq}{v_F}+\frac{16\alpha c}{v_F^2}k_BT\ln 2,
\label{eq13}
\end{equation}
\noindent
where $\alpha=e^2/(\hbar c)$ is the fine structure constant and $v_F\approx c/300$
is the Fermi velocity of graphene.

Using Eqs.~(\ref{eq4}), (\ref{eq12}) and (\ref{eq13}), we find
\begin{equation}
q_{\mu}\sigma^{{\rm K},\mu 0}(\omega,\mbox{\boldmath$q$})=
-i\frac{\alpha c^2}{4\pi\hbar v_F}\left(\pi\hbar q+
\frac{16k_B T}{v_F}\ln 2\right)\neq 0,
\label{eq14}
\end{equation}
\noindent
i.e., for the modified expressions for the
polarization and conductivity used in Ref.~\cite{1} the
transversality condition for $\nu=0$ is violated.
In fact the quantities $\widetilde{\Pi}_{\mu\nu}$ and
 $\sigma_{\mu\nu}^{\rm K}$ are not tensors because the subtraction made in
 Eq.~(\ref{eq4}) violated a tensorial structure.

It has been known that the $i,\,j=1,\,2$
 components of the polarization and
conductivity tensors can be presented in terms of the longitudinal and
transverse quantities \cite{22}
\begin{eqnarray}
&&
\Pi_{ij}(\omega,\mbox{\boldmath$q$})=\frac{q_iq_j}{q^2}
\Pi_L(\omega,\mbox{\boldmath$q$}) +\left(\delta_{ij}-\frac{q_iq_j}{q^2}\right)
\Pi_T(\omega,\mbox{\boldmath$q$}),
\nonumber\\
&&
\sigma_{ij}(\omega,\mbox{\boldmath$q$})=\frac{q_iq_j}{q^2}
\sigma_L(\omega,\mbox{\boldmath$q$}) +\left(\delta_{ij}-\frac{q_iq_j}{q^2}\right)
\sigma_T(\omega,\mbox{\boldmath$q$}).
\label{eq15}
\end{eqnarray}

We set the components of the first equality in Eq.~(\ref{eq15}) equal to
the corresponding components in Eq.~(\ref{eq2}) and obtain
\begin{equation}
\Pi_L(\omega,\mbox{\boldmath$q$})=\frac{\omega^2}{c^2q^2}
\Pi_{00}(\omega,\mbox{\boldmath$q$}), \quad
\Pi_T(\omega,\mbox{\boldmath$q$})=-\frac{1}{q^2}
\Pi(\omega,\mbox{\boldmath$q$}).
\label{eq16}
\end{equation}
\noindent
In a similar way, using the conductivity tensor (\ref{eq3}) and the second
equality in Eq.~(\ref{eq15}), one finds
\begin{equation}
\sigma_L(\omega,\mbox{\boldmath$q$})=\frac{\omega^2}{c^2q^2}
\sigma_{00}(\omega,\mbox{\boldmath$q$}), \quad
\sigma_T(\omega,\mbox{\boldmath$q$})=-\frac{1}{q^2}
\sigma(\omega,\mbox{\boldmath$q$}),
\label{eq17}
\end{equation}
\noindent
where
\begin{equation}
\sigma(\omega,\mbox{\boldmath$q$})=q^2
\sigma_{\mu}^{\,\mu}(\omega,\mbox{\boldmath$q$})+
\left(\frac{\omega^2}{c^2}-q^2\right)
\sigma_{00}(\omega,\mbox{\boldmath$q$})
=\frac{c^2}{4\pi\hbar i\omega}
\Pi(\omega,\mbox{\boldmath$q$}).
\label{eq18}
\end{equation}

By representing the modified expressions for the
polarization and conductivity (\ref{eq4})
used in Ref.~\cite{1} in the form of Eq.~(\ref{eq15}) and repeating the same
calculations with the help of Eqs.~(\ref{eq2}) and (\ref{eq5}), we arrive at
\begin{eqnarray}
&&
\widetilde{\Pi}_L(\omega,\mbox{\boldmath$q$})=\frac{\omega^2}{c^2q^2}
\Pi_{00}(\omega,\mbox{\boldmath$q$})=
\Pi_L(\omega,\mbox{\boldmath$q$}),
\nonumber\\
&&
\widetilde{\Pi}_T(\omega,\mbox{\boldmath$q$})=-\frac{1}{q^2}
\left[\Pi(\omega,\mbox{\boldmath$q$})-
\lim\limits_{\omega\to 0}\Pi(\omega,\mbox{\boldmath$q$})\right]
\label{eq19}
\end{eqnarray}
\noindent
and
\begin{eqnarray}
&&
{\sigma}_L^{\rm K}(\omega,\mbox{\boldmath$q$})=\frac{\omega^2}{c^2q^2}
\sigma_{00}(\omega,\mbox{\boldmath$q$})=
\sigma_L(\omega,\mbox{\boldmath$q$}),
\nonumber\\
&&
\sigma_T^{\rm K}(\omega,\mbox{\boldmath$q$})=-\frac{1}{q^2}
\left[\sigma(\omega,\mbox{\boldmath$q$})-
\lim\limits_{\omega\to 0}\sigma(\omega,\mbox{\boldmath$q$})\right]
\nonumber\\
&&
\phantom{\sigma_T^{\rm K}(\omega,\mbox{\boldmath$q$})}
=\frac{ic^2}{4\pi\hbar\omega q^2}
\left[\Pi(\omega,\mbox{\boldmath$q$})-
\lim\limits_{\omega\to 0}\Pi(\omega,\mbox{\boldmath$q$})\right].
\label{eq20}
\end{eqnarray}

As correctly concluded in Ref.~\cite{1}, the made modification (\ref{eq4})
does not change the longitudinal conductivity of graphene and modifies only the
transverse one. Below we focus on the transverse part and
demonstrate that the results obtained using this modification  are physically unacceptable.

Thus, for a pristine graphene at $T=0$ the quantity $\Pi$ is given by
\cite{2,3,4,5,13,14}
\begin{equation}
\Pi(\omega,\mbox{\boldmath$q$})=\frac{\pi e^2q^2}{c^2}\left\{
\begin{array}{ll}
\phantom{\mp i}\sqrt{v_F^2q^2-\omega^2}, & |\omega|<v_Fq, \\[3mm]
\mp i\sqrt{\omega^2-v_F^2q^2}, & |\omega|>v_Fq,
\end{array}
\right.
\label{eq21}
\end{equation}
\noindent
where here and below the upper and lower signs are for $\omega >v_Fq$ and
$\omega <-v_Fq$, respectively.

{}From Eq.~(\ref{eq21}) at fixed $q$ one finds
\begin{equation}
\lim\limits_{\omega\to 0}\Pi(\omega,\mbox{\boldmath$q$})=
\frac{\pi e^2v_Fq^3}{c^2}.
\label{eq22}
\end{equation}
\noindent
This result is valid not only at $T=0$ but at any temperature \cite{21}.
Using Eqs.~(\ref{eq20})-(\ref{eq22}), for the modified transverse
conductivity we obtain
\begin{equation}
\sigma_T^{\rm K}(\omega,\mbox{\boldmath$q$})=\frac{\sigma_0}{\omega}\left\{
\begin{array}{ll}
i\left(\!\sqrt{v_F^2q^2-\omega^2}-v_Fq\right), & |\omega|<v_Fq, \\[3mm]
\pm \sqrt{\omega^2-v_F^2q^2}-iv_Fq, & |\omega|>v_Fq,
\end{array}
\right.
\label{eq23}
\end{equation}
\noindent
where $\sigma_0=e^2/(4\hbar)$ is the universal conductivity of graphene.
Note that the regions $0<\omega<v_Fq$ and $v_Fq<\omega<cq$ correspond to the
evanescent waves whereas for the
propagating waves $0\leqslant cq\leqslant\omega$ holds.

As is seen in Eq.~(\ref{eq23}), for $\omega\geqslant cq$ Ref.~\cite{1}
arrives at
\begin{equation}
{\rm Im}\,\sigma_T^{\rm K}(\omega,\mbox{\boldmath$q$})=-\frac{\sigma_0v_Fq}{\omega}
\label{eq24}
\end{equation}
\noindent
in  contradiction with the previously obtained conclusion that at zero temperature
the conductivity of pure graphene in the region of propagating waves is real
at all frequencies
\cite{23,24}. The result following from the quantum field theoretical definition
(\ref{eq3}) of the conductivity tensor
\begin{equation}
\sigma_T(\omega,\mbox{\boldmath$q$})=\frac{\sigma_0}{\omega}\left\{
\begin{array}{ll}
i\sqrt{v_F^2q^2-\omega^2}, & |\omega|<v_Fq, \\[3mm]
\pm \sqrt{\omega^2-v_F^2q^2}, & |\omega|>v_Fq
\end{array}
\right.
\label{eq25}
\end{equation}
\noindent
is in agreement with this conclusion.

According to Ref.~\cite{1}, the relation between the polarization tensor and
the electric current
\begin{equation}
J_{\mu}(\omega,\mbox{\boldmath$q$})=\frac{c}{4\pi \hbar}
\Pi_{\mu\nu}(\omega,\mbox{\boldmath$q$})
A^{\nu}(\omega,\mbox{\boldmath$q$})
\label{eq26}
\end{equation}
\noindent
used in Refs.~\cite{10,12,15,16,19} for a derivation  of Eq.~(\ref{eq3})
is unsatisfactory because it leads to a nonzero electric current expressed
via $\lim\limits_{\omega\to 0}\Pi_{\mu\nu}(\omega,\mbox{\boldmath$q$})$
for the equal to zero electric field. Based on this, Ref.~\cite{1}
replaces $\Pi_{\mu\nu}$ in Eq.~(\ref{eq26}) with the modified expression
$\widetilde{\Pi}_{\mu\nu}$ defined in Eq.~(\ref{eq4}). Below we show that this
replacement has no justification.

Really, Ref.~\cite{1} uses the temporal gauge and considers
the zero electric field,
$E^{\nu}(t,\mbox{\boldmath$q$})=0$, as the negative derivative with respect
to $t$ of the constant in time vector potential $A_0^{\,\nu}(\mbox{\boldmath$q$})$
with fixed {\boldmath$q$}.
It is true that in the region of evanescent waves, $0<\omega<v_Fq$, where it is
possible to consider limit $\omega\to 0$ at fixed $q$, using Eqs. (\ref{eq2}) and
(\ref{eq22}), one obtains
\begin{equation}
\lim\limits_{\omega\to 0}\Pi_{ii}(\omega,\mbox{\boldmath$q$})=-\left(1-
\frac{q_i^2}{q^2}\right)\frac{\pi e^2v_Fq}{c^2}\neq 0.
\label{eq22new}
\end{equation}

Thus, in the region of evanescent waves, at any temperature, including $T=0$, the transverse conductivity $\sigma_T$ found in the framework of quantum field theory
has an imaginary part. For instance, at $T=0$, where it is pure imaginary, using
 the first line of Eq.~(\ref{eq21}), we obtain
\begin{equation}
\sigma_{T}(\omega,\mbox{\boldmath$q$})=i
\sigma_0\frac{\sqrt{v_F^2q^2-\omega^2}}{\omega}.
\label{eq38}
\end{equation}

According to Ref.~\cite{1}, Eq.~(\ref{eq22new}) results in the appearance of an
electric current in graphene in the absence of electric field, whereas
Eq.~(\ref{eq38}) leads to an unacceptable
double pole at zero frequency in the graphene dielectric permittivity.
If the expression (\ref{eq4}) is used \cite{1}, at zero temperature
the modified polarization,
 $\widetilde{\Pi}_{\mu\nu}(\omega,\mbox{\boldmath$q$})$, and
 conductivity, $\sigma_{T}^{\rm K}(\omega,\mbox{\boldmath$q$})$,
expressions  vanish in the limit of zero frequency.

Note, however, that at nonzero temperature in the region of propagating waves the
pure imaginary current in the absence of electric field arises both in the
quantum field theoretical formalism
using Eq.~(\ref{eq26}) and in the approach
of Ref.~\cite{1} which replaces the polarization tensor $\Pi_{\mu\nu}$ in
Eq.~(\ref{eq26}) with the modified  expression,
$\widetilde{\Pi}_{\mu\nu}$, defined in
Eq.~(\ref{eq4}).

Really, for the propagating waves the condition $0\leqslant cq\leqslant\omega$
holds. Then, if $\omega\to 0$, the wave vector $q$ must go to zero as well.
For the real parts of $\Pi_{00}$ and $\Pi$,
under the condition $\hbar\omega\ll k_BT$, one has \cite{2}
\begin{eqnarray}
&&
{\rm Re}\,\Pi_{00}(\omega,\mbox{\boldmath$q$})=-8\ln 2\frac{e^2k_BT}{\hbar}\,
\frac{q^2}{\omega^2}\left[1+O\left(\frac{v_F^2q^2}{\omega^2}\right)\right],
\nonumber \\
&&
{\rm Re}\,\Pi(\omega,\mbox{\boldmath$q$})=8\ln 2\frac{e^2k_BT}{\hbar}\,
\frac{q^2}{c^2}\left[1+O\left(\frac{v_F^2q^2}{\omega^2}\right)\right].
\label{eq31}
\end{eqnarray}

With the help of these results, using Eq.~(\ref{eq2}), we find
\begin{equation}
{\rm Re}\,\Pi_{11}(\omega,\mbox{\boldmath$q$})=
{\rm Re}\,\Pi_{22}(\omega,\mbox{\boldmath$q$})=
-8\ln 2\frac{e^2k_BT}{\hbar c^2},
\label{eq32}
\end{equation}
\noindent
i.e., these components of the polarization tensor in the lowest order with
respect to the small parameter $v_F^2q^2/\omega^2$ are equal to the constant
independent on $\omega$ and $q$. Thus, they preserve their value (\ref{eq32})
in the limit $\omega,\,q\to 0$.

Using Eqs.~(\ref{eq16}) and (\ref{eq31}), we also obtain
\begin{equation}
{\rm Re}\,\Pi_{L}(\omega,\mbox{\boldmath$q$})=
{\rm Re}\,\Pi_{T}(\omega,\mbox{\boldmath$q$})=
-8\ln 2\frac{e^2k_BT}{\hbar c^2}.
\label{eq33}
\end{equation}

The corresponding values of the longitudinal and transverse conductivities are
obtained from Eqs.~(\ref{eq3}), (\ref{eq17}), (\ref{eq18}), and (\ref{eq31})
\begin{equation}
{\rm Im}\,\sigma_{L}(\omega,\mbox{\boldmath$q$})=
{\rm Im}\,\sigma_{T}(\omega,\mbox{\boldmath$q$})=
\sigma_0\frac{8\ln 2}{\pi}\,\frac{k_BT}{\hbar\omega}.
\label{eq34}
\end{equation}
\noindent
Note that in the region of propagating waves the result (\ref{eq34})
leads to a simple pole in the corresponding dielectric permittivities
of graphene
\begin{eqnarray}
&&
{\rm Re}\,\varepsilon_{{\rm L,T}}(\omega,\mbox{\boldmath$q$})=
1-\frac{2\pi q}{\omega}{\rm Im}\,\sigma_{{\rm L,T}}(\omega,\mbox{\boldmath$q$})
\nonumber \\
&&~~
=1-16\ln 2\frac{\sigma_0}{c}\,\sin\theta\,\frac{k_BT}{\hbar\omega},
\label{eq26a}
\end{eqnarray}
\noindent
where $\theta$ is the angle of incidence of the wave on a graphene sheet,
$\sin\theta=cq/\omega=\mbox{const}$.

For the modified expressions used in Ref.~\cite{1} from Eqs.~(\ref{eq19}),
(\ref{eq22}), and (\ref{eq31}) one finds
\begin{eqnarray}
&&
{\rm Re}\,\widetilde{\Pi}_{L}(\omega,\mbox{\boldmath$q$})=
{\rm Re}\,\Pi_{L}(\omega,\mbox{\boldmath$q$}),
\nonumber \\
&&
{\rm Re}\,\widetilde{\Pi}_T(\omega,\mbox{\boldmath$q$})=-8\ln 2\frac{e^2k_BT}{\hbar c^2}\,
+\frac{\pi e^2v_Fq}{c^2}.
\label{eq35}
\end{eqnarray}

The corresponding conductivities found from Eq.~(\ref{eq20}) have the imaginary parts
\begin{eqnarray}
&&
{\rm Im}\,\sigma_{L}^{\rm K}(\omega,\mbox{\boldmath$q$})=
\sigma_0\frac{8\ln 2}{\pi}\,\frac{k_BT}{\hbar\omega},
\nonumber \\
&&
{\rm Im}\,\sigma_{T}^{\rm K}(\omega,\mbox{\boldmath$q$})=
\sigma_0\left(\frac{8\ln 2}{\pi}\,\frac{k_BT}{\hbar\omega}-
\frac{v_Fq}{\omega}\right).
\label{eq36}
\end{eqnarray}

What is important, the diagonal components of the modified polarization
expression found from Eq.~(\ref{eq15}) with added tildes using Eq.~(\ref{eq35})
do no  vanish in the limit $\omega,\,q\to 0$
\begin{equation}
\lim\limits_{\omega,\,q\to 0}{\rm Re}\,\widetilde{\Pi}_{11}(\omega,\mbox{\boldmath$q$})=
\lim\limits_{\omega,\,q\to 0}{\rm Re}\,\widetilde{\Pi}_{22}(\omega,\mbox{\boldmath$q$})=
-8\ln 2\frac{e^2k_BT}{\hbar c^2}.
\label{eq37}
\end{equation}

Thus, at nonzero temperature in both the quantum field theory and in the formalism
of Ref.~\cite{1}  a nonzero current in graphene arises even for a zero electric field.
Keeping in mind, however, that the conductivity in both cases is pure imaginary
because its real part vanishes when $\omega,\,q\to 0$ \cite{2},
this creates no problem. Such a pure imaginary current satisfies the relativistically
covariant Ohm's law \cite{24a}
\begin{equation}
J_{\mu}(\omega,\mbox{\boldmath$q$})=
\sigma_{\mu\nu}(\omega,\mbox{\boldmath$q$})\,E^{\nu},
\label{eq33a}
\end{equation}
\noindent
where $\sigma_{\mu\nu}$ goes to $i\infty$ with vanishing $E^{\nu}$.
Note that according to Ref.~\cite{1} the current in Eq.~(\ref{eq33a}) is different
from that in Eq.~(\ref{eq26}) because the latter contains the magnetically induced
part. This statement is, however, incorrect. In the absence of the
constant in time, external magnetic
field, both Eqs.~(\ref{eq26}) and (\ref{eq33a}) describe the total induced
current \cite{19a}.

It seems illogical that the same behavior of $\sigma_L^{\rm K}$ and $\sigma_T^{\rm K}$
in Eq.~(\ref{eq36}) following from the modified formalism of Ref.~\cite{1}
in the region of propagating waves as of $\sigma_L$ and $\sigma_T$ in Eqs.~(\ref{eq38})
 and (\ref{eq34}) using quantum field theory is not considered in Ref.~\cite{1}
 as leading to a permanent
electric current in the absence of electric field.
It should be noted that Ref.~\cite{1} introduces the phenomenological parameter
$\Gamma$ describing the dissipation of electronic quasiparticles by replacing the
frequency $\omega$ with $\omega+i\Gamma$ (see below). At the cost of this replacement,
the pure imaginary currents arising in the absence of nonzero electric field can be
eliminated. However, an incorporation of the dissipation of electronic quasiparticles
brings the approach of Ref.~\cite{1} outside the application region of the Dirac model,
where the quasiparticles are considered as noninteracting.
Note also that just the behavior
of the conductivities of graphene according to Eq.~(\ref{eq34}) leads to the big thermal effect in the Casimir force between two graphene sheets at
short separations predicted in Ref.~\cite{25} for the case of two pristine
graphene sheets and confirmed experimentally in
Refs.~\cite{26,27}.

Hence, the modified expression (\ref{eq4}) was obtained in Ref.~\cite{1}
by inappropriately using the nonrelativistic realization of the causality
condition in the relativistic Dirac model. 
This modification caused a violation
of the condition of gauge invariance and the physical inconsistencies described
above. As to a prediction of the double pole at zero frequency by
quantum field theory, it does not contradict to any physical
results, including the Kubo approach formulated for the propagating fields, and
suggests a path to the resolution of long-term problems in the Casimir effect \cite{13}.

Last but not least,
Ref.~\cite{1} does not inform the reader about the used system of units.
Thus, in Table~I on p.3 $k_0=\omega$ (left column) and $k=k_{\mu}=(k_0,\tilde{k}_{||})$
where $k_0$ has the dimension of s$^{-1}$ and $\tilde{k}_{||}$ has the dimension of energy
(many equations of Ref.~\cite{1} contain $c$ and $\hbar$ which are not put to unity).
As is stated in Ref.~\cite{1}, the polarization tensor  used there is ``exactly identical" to that
derived within the quantum field theoretical approach. However, according to Eqs.~(159)
and (160), the polarization tensor of Ref.~\cite{1} has the dimension of cm/s$^2$,
whereas the polarization tensor of Refs.~\cite{2,3,4} has the dimension of g\,cm/s.

What is more, the expression connecting the polarization and conductivity tensors is
written in the form used in Refs.~\cite{10,12} where the system of units with
$\hbar=c=k_B=\varepsilon_0=1$ has been used ($\varepsilon_0$ is the dielectric
permittivity of vacuum).  In contradiction with this, Ref.~\cite{1} uses the
expression for the
fine structure constant $\alpha=e^2/(\hbar c)$, whereas in Refs.~\cite{10,12} the
expression $\alpha=e^2/(4\pi)$ has been used. The definition of $q_z$ below Eq.~(120)
is in contradiction with the definition of the same quantity given in TABLE I, right
column, line 9.
Above Eq.~(C34) Ref.~\cite{1} considers ``the case $\Pi_{\mu\nu}(q)=\Pi_L(q)$" which is
meaningless if to take into account the expression for $\Pi_L(q)$ in Eq.~(121).

To conclude, in the foregoing it was shown that the modification of the polarization
and conductivity tensors made in Ref.~\cite{1} violates their transversality and
leads to other physically unacceptable
consequences. This modification is superfluous because the claims of Ref.~\cite{1} about
the problems arising when using the standard
formalism of quantum field theory are invalid.
Equation (\ref{eq4}) was  derived from the Kubo formula by using the nonrelativistic
 realization of causality in application to the relativistic Dirac model
where it is inapplicable.

It should be added also that, according to Ref.~\cite{1}, the quantum field theoretical
description of the conductivity of graphene does not describe unavoidable losses.
This is, however, not the case because the conductivity of graphene obtained using this
description possesses both the real and imaginary parts. In doing so, the real part
of conductivity results in the positive imaginary part of the dielectric permittivity of
graphene which describes the losses of conduction electrons in graphene.
As to the massless or very light electronic quasiparticles, in the application region
of the Dirac model they are considered as noninteracting. By introducing the
phenomenological parameter $\Gamma$ for a description the relaxation of
quasiparticles in graphene, Ref.~\cite{1} again violated the relativistic
description of graphene inherent to the Dirac model.

All the above examples were given for  the case of a pristine graphene
in order to avoid unnecessary technical difficulties and to make the presentation
maximally transparent. However, in the application
region of Dirac model, all the statements made remain valid for a general
case of graphene possessing some nonzero energy
gap and chemical potential.

The authors are grateful to D. V. Vassilevich for useful comments.
The work of G.L.K.\ and V.M.M.~was supported by the State Assignment for Basic Research
(project FSEG-2023-0016). N.K. was supported in part by the grants 2025/13673-0,
2021/10128-0 of S{\~a}o Paulo Research Foundation (FAPESP).



\begin{thebibliography}{99}
\bibitem{1}
P.~Rodriguez-Lopez, J.-S.~Wang, and M.~Antezza,
Electric conductivity in graphene: Kubo model versus a nonlocal quantum
field theory model,
{Phys. Rev.} B {\bf 111}, 115428 (2025).
\bibitem{2}
G.~L.~Klimchitskaya and V.~M.~Mostepanenko,
Conductivity of pure graphene: Theoretical approach using the polarization tensor,
{Phys. Rev.} B {\bf 93}, 245419 (2016).
\bibitem{3}
G.~L.~Klimchitskaya and V.~M.~Mostepanenko,
Quantum electrodynamic approach to the conductivity of gapped graphene,
{Phys. Rev.} B {\bf 94}, 195405 (2016).
\bibitem{4}
G.~L.~Klimchitskaya, V.~M.~Mostepanenko, and V.~M.~Petrov,
Conductivity of graphene in the framework of Dirac model:
Interplay between nonzero mass gap and chemical potential,
{Phys. Rev.} B {\bf 96}, 235432 (2017).
\bibitem{5}
G.~L.~Klimchitskaya and V.~M.~Mostepanenko,
Kramers-Kronig relations and causality conditions for graphene in the
framework of the Dirac model,
Phys. Rev. D {\bf 97}, 085001 (2018).
\bibitem{6}
M.~Bordag, I.~V.~Fialkovsky, D.~M.~Gitman, and
D.~V.~Vassilevich,
Casimir interaction between a perfect conductor and graphene
described by the Dirac model,
{Phys. Rev. B} {\bf 80}, 245406 (2009).
\bibitem{7}
I.~V.~Fialkovsky, V.~N.~Marachevsky, and
D.~V.~Vassilevich,
Finite-temperature Casimir effect for graphene,
{Phys. Rev. B} {\bf 84}, 035446 (2011).
\bibitem{8}
M.~Bordag, G.~L.~Klimchitskaya, V.~M.~Mostepanenko, and V.~M.~Petrov,
Quantum field theoretical description for the reflectivity of graphene,
Phys. Rev. D {\bf 91}, 045037 (2015); {\bf 93}, 089907(E) (2016).
\bibitem{9}
M.~Bordag, I.~Fialkovskiy, and D.~Vassilevich,
Enhanced Casimir effect for doped graphene,
Phys. Rev. B {\bf 93}, 075414 (2016);
{\bf 95}, 119905(E) (2017).
\bibitem{10}
I.~V~Fialkovskiy and D.~V.~Vassilevich,
Quantum field theory in graphene,
Int. J. Mod. Phys. A {\bf 27}, 1260007 (2012).
\bibitem{12}
I.~V.~Fialkovskiy and D.~V.~Vassilevich,
Graphene through the looking glass of QFT,
 Mod. Phys. Lett. A {\bf 31}, 1630047 (2016).
\bibitem{13}
G.~L.~Klimchitskaya and V.~M.~Mostepanenko,
Quantum field theoretical framework for the electromagnetic response of
graphene and dispersion relations with implications to the Casimir effect,
Phys. Rev. D {\bf 107}, 105007 (2023).
\bibitem{14}
M.~Bordag, G.~L.~Klimchitskaya, and V.~M.~Mostepanenko,
Convergence of the polarization tensor in spacetime of three dimensions,
Phys. Rev. D {\bf 109}, 125014 (2024).
\bibitem{15}
N.~Khusnutdinov and D.~V.~Vassilevich,
Impurities in graphene and their influence on the Casimir interaction,
Phys. Rev. B {\bf 109}, 235420 (2024).
\bibitem{16}
N.~Khusnutdinov and N.~Emelianova,
The polarization tensor approach for Casimir effect,
Int. J. Mod. Phys. A {\bf 40}, 2543004 (2025).
\bibitem{17}
P.~K.~Pyatkovsky,
Dynamic polarization, screening, and plasmons in gapped graphene,
J. Phys.: Condens. Matter {\bf 21}, 025506 (2009).
\bibitem{18}
G.~L.~Klimchitskaya and V.~M.~Mostepanenko,
Casimir and Casimir-Polder Forces in Graphene
Systems: Quantum Field Theoretical Description
and Thermodynamics,
Universe {\bf 6}, 150 (2020).
\bibitem{19}
V.~P.~Gusynin, S.~G.~Sharapov, and J.~P.~Carbotte,
Magnetooptical conductivity in graphene,
J. Phys.: Condens. Matter {\bf 19}, 026222 (2007).
\bibitem{19a}
D.~B.~Melrose,
{\it Quantum Plasmadynamics: Unmagnetized Plasmas}
(Springer, New York, 2008).
\bibitem{19b}
H.~Bruus and K.~Flensberg,
{\it Many-Body Quantum Theory in Condensed Matter Physics:
An Introduction}
(Oxford University Press, Oxford, New York, 2016).
\bibitem{19c}
V.~P.~Gusynin and S.~G.~Sharapov,
Transport of Dirac quasiparticles in graphene: Hall and optical conductivities
{Phys. Rev. B} {\bf 73}, 245411 (2006).
\bibitem{19d}
V.~P.~Gusynin, S.~G.~Sharapov, and J.~P.~Carbotte,
AC conductivity of graphene: From tight-binding model to 2 + 1-
dimensional quantum electrodynamics,
{Int. J. Mod. Phys. B} {\bf 21}, 4611--4658 (2007).
\bibitem{20}
L.~D.~Landau and E.~M.~Lifshitz,
{\it The Classical Theory of Fields}
(Elsevier, Amsterdam, 1980).
\bibitem{21}
G.~L.~Klimchitskaya, C.~C.~Korikov, and V.~M.~Mostepanenko,
Polarization tensor in spacetime of three dimensions and a quantum
field-theoretical description of the nonequilibrium Casimir force
in graphene systems,
Phys. Rev. A {\bf 111}, 012812 (2025).
\bibitem{22}
L.~D.~Landau, E.~M.~Lifshitz, and  L.~P.~Pitaevskii, L.P.
{\it Electrodynamics of Continuous Media}
(Pergamon, Oxford, 1984).
\bibitem{23}
L.~A.~Falkovsky and A.~A.~Varlamov,
Space-time dispersion of graphene conductivity,
Eur. Phys. J. B {\bf 56}, 281 (2007).
\bibitem{24}
M.~I.~Katsnelson,
{\it The Physics of Graphene}
(Cambridge University Press, Cambridge, 2020).
\bibitem{24a}
R. Starke and G. A. H. Schober,
Relativistic covariance of Ohm's law,
{Int. J. Mod. Phys. D} {\bf 25}, 1640010 (2016).
\bibitem{25}
G.~G\'{o}mez-Santos,
Thermal van der Waals interaction between graphene layers,
Phys. Rev. B {\bf 80}, 245424 (2009).
\bibitem{26}
M.~Liu, Y.~Zhang, G.~L.~Klimchitskaya, V.~M.~Mostepanenko, and U.~Mohideen,
Demonstration of an Unusual Thermal Effect in the Casimir Force from
Graphene,
Phys. Rev. Lett. {\bf 126}, 206802 (2021).
\bibitem{27}
M.~Liu, Y.~Zhang, G.~L.~Klimchitskaya, V.~M.~Mostepanenko, and U.~Mohideen,
Experimental and theoretical investigation of the thermal effect in the
Casimir interaction from graphene,
Phys. Rev. B {\bf 104}, 085436 (2021).
\end{thebibliography}
\end{document}